\documentclass[reprint,amsmath,amssymb,aps,prd,superscriptaddress]{revtex4-2}
\usepackage[utf8]{inputenc}
\usepackage{graphicx,amssymb,amstext,amsmath}
\usepackage{dcolumn}
\usepackage{multirow}
\usepackage{natbib}
\usepackage{xspace}

\newcommand{\GeV}{\ensuremath{\,\text{Ge\hspace{-.08em}V}}\xspace}

\newcommand{\pt}{\ensuremath{p_{\mathrm{T}}}\xspace}
\newcommand{\pT}{\ensuremath{p_{\mathrm{T}}}\xspace}

\newcommand{\PTm}{\ensuremath{{p}_\mathrm{T}\hspace{-1.02em}/\kern 0.5em}\xspace}

\newcommand{\ETslash}{\ensuremath{E_{\mathrm{T}}\hspace{-1.1em}/\kern0.45em}\xspace}

\newcommand{\ptvecmiss}{\ensuremath{{\vec p}_{\mathrm{T}}^{\kern1pt\text{miss}}}\xspace}

\newcommand{\cmsSymbolFace}{\mathrm}

\newcommand{\ttbar}{\ensuremath{\cmsSymbolFace{t}\overline{\cmsSymbolFace{t}}}\xspace} 

\newcommand{\rpv}{\ensuremath{\rlap{\kern.2em/}R}\xspace}

\begin{document}

\title{End-to-End Jet Classification of Boosted Top Quarks\\ with the CMS Open Data}

\author{M.~Andrews}%
    \affiliation{Department of Physics, Carnegie Mellon University, Pittsburgh, Pennsylvania 15213, USA}
\author{B.~Burkle}%
    \affiliation{Department of Physics, Brown University, Providence, Rhode Island 02912, USA}
\author{Y.~Chen}\affiliation{Google Inc., Mountain View, California 94043, USA}
\author{D.~DiCroce}\affiliation{Department of Physics and Astronomy, University of Alabama, Tuscaloosa, Alabama 35487, USA}
\author{S.~Gleyzer}\affiliation{Department of Physics and Astronomy, University of Alabama, Tuscaloosa, Alabama 35487, USA}
\author{U.~Heintz}\affiliation{Department of Physics, Brown University, Providence, Rhode Island 02912, USA}
\author{M.~Narain}\affiliation{Department of Physics, Brown University, Providence, Rhode Island 02912, USA}
\author{M.~Paulini}\affiliation{Department of Physics, Carnegie Mellon University, Pittsburgh, Pennsylvania 15213, USA}
\author{N.~Pervan}\affiliation{Department of Physics, Brown University, Providence, Rhode Island 02912, USA}
\author{Y.~Shafi}\affiliation{Google Inc., Mountain View, California 94043, USA}
\author{W.~Sun}\affiliation{Google Inc., Mountain View, California 94043, USA}
\author{E.~Usai}\affiliation{Department of Physics, Brown University, Providence, Rhode Island 02912, USA}
\author{K.~Yang}\affiliation{Google Inc., Mountain View, California 94043, USA}

\date{\today}

\begin{abstract}
We describe a novel application of the end-to-end deep learning technique to the task of discriminating top quark-initiated jets from those originating from the hadronization of a light quark or a gluon.
The end-to-end deep learning technique uses low-level detector representation of high-energy collision event as inputs to deep learning algorithms.
In this study, we use low-level detector information from the simulated CMS Open Data samples to construct the top jet classifiers.
To optimize classifier performance we progressively add low-level information from the CMS tracking detector, including pixel detector reconstructed hits and impact parameters, and demonstrate the value of additional tracking information even when no new spatial structures are added.
Relying only on calorimeter energy deposits and reconstructed pixel detector hits, the end-to-end classifier achieves an AUC score of 0.975$\pm$0.002 for the task of classifying boosted top quark jets.
After adding derived track quantities, the classifier AUC score increases to 0.9824$\pm$0.0013, serving as the first performance benchmark for these CMS Open Data samples.
 
\end{abstract}

\maketitle

\section{Introduction\label{sec:Introduction}}
The Large Hadron Collider (LHC) is a prolific top quark factory: since the beginning of data-taking in 2010, over $10^8$ top quarks have been produced. The measurement of the top quark's properties and production rates at the LHC remains one of the main research priorities at experiments like the Compact Muon Solenoid (CMS) at the LHC.
Moreover, investigating high transverse moment top quark production offers potential hints of the presence of new physics that may lie beyond the Standard Model.

Top quarks are unique in that they decay before they have time to hadronize, always decaying to a bottom quark and a W-boson. During the top decay chain, the W-boson will decay hadronically to quarks 66.5\% or leptonically to a lepton and neutrino pair 33.5\% of the time \cite{pdg}. At hadron colliders, like the LHC, the low production cross section of prompt electrons and muons can be exploited to boost tagging efficiency when identifying top quarks with a leptonically decaying W-boson in its decay chain. However, hadronic decays of top quarks can be much harder to identify, since the primary features used to identify them are the topology of its decay products and the track features of the bottom quark decay products.
In particular, at high transverse momenta, the hadronic decay of highly a Lorentz-boosted top quark can lead to a single merged cluster of particles in the detector, hereby referred to as jets, offering a unique and challenging view into the study of the top quark's properties. Because of this, discriminating boosted top quark-jets from light flavour- or gluon-jets has become an important challenge for the LHC experiments, and a popular benchmark for data analysis techniques involving machine learning. 

Most jet identification techniques rely on inputs provided by the Particle Flow (PF) algorithm used to convert detector level information to physics objects \cite{PF}. The Particle Flow algorithm has many advantages due to its ability to greatly reduce the size and complexity of particle physics data while providing a physically intuitive and easy to use representation in physics analyses. 
Many of the modern machine learning approaches to jet discrimination are based on PF-based inputs 
\cite{deepak8,htt,cmstt,energyflow,mother,best,atlas,jmar}.
However, there is some invariable loss of information from reducing the data set complexity. Despite the very high reconstruction efficiency of PF algorithms, some physics objects may fail to be reconstructed, are reconstructed imperfectly, or exist as fakes \cite{tdr-performance}. For that reason it is advantageous to consider end-to-end reconstruction that allows a direct application of machine learning algorithms to low-level data representation in the detector.

In this work, we extend the end-to-end deep learning approach for particle and event classification \cite{e2e1}. Specifically, we extend the use of end-to-end jet images introduced for quark- vs. gluon-jet discrimination \cite{e2e2} to the task of boosted top quark- vs. light quark- or gluon-jet discrimination. In previous work \cite{e2e2}, we found that the track information was the leading contributor to the classifier's performance. Due to this insight and the importance of identifying displaced tracks associated with bottom quark decays, this new work introduces a number of key features from the CMS tracking detectors to exploit the full topology of hadronically decaying top quarks.

\section{Open Data Simulated Samples\label{sec:Samples}}

The end-to-end deep learning technique relies on high-fidelity simulated detector data, which in this work comes from the simulated Monte Carlo in the CMS Open Data Portal \cite{opendataweb, opendatacms}. We use a sample of SM top-antitop ($t\bar{t}$) pair production as a source of boosted top quarks, where the $W$ boson from the top quark decay is required to decay to quarks \cite{ttbarod}. Additionally, the reconstructed top quark transverse momentum ($p_{T}$) is required to be greater than 400$\GeV$. At this momentum, we expect that a large fraction of the $W$- and b-jets produced in the top quark decay chain will be suitably merged. The Monte Carlo sample was generated with Madgraph 2.6.6 \cite{madgraph} and uses Pythia6 for parton showering \cite{pythia6} with the Z2Star tune. For the light-flavour and gluon jets, we use three samples of QCD dijet production in different ranges of the hard-scatter transverse momentum ($\hat{p}_{T}$): $300<\hat{p}_{T}<600\GeV$, $400<\hat{p}_{T}<600\GeV$, and  $600<\hat{p}_{T}<3000\GeV$ \cite{qcdod1, qcdod2, qcdod3}. 
Like the $\ttbar$ sample, these samples were generated and showered with Pythia6 \cite{pythia6} using the same Z2Star tune.
For all samples, the detector response is simulated using Geant4 with the full CMS geometry and is processed through the CMS PF reconstruction algorithm using CMSSW release 5\_3\_32 \cite{cmssw}. An average of ten additional background collisions or pileup (PU) interactions are added to the simulated hard-scatter event, which are sampled from a realistic distribution of simulated minimum bias events.
For this study, we use a custom CMS data format which includes the low-level tracker detector information, specifically, the reconstructed clusters from the pixel and silicon strip detectors \cite{opendataml}. From the tracker clusters, we perform a parametric estimate of the position of the hit on the sensor surface.

\begin{table}[htbp] 
\centering
\begin{tabular}{l c c | c}
\hline\hline
\textbf{Category} & \textbf{ Top quark jets } & \textbf{ QCD jets }  & \textbf{ Total Jets }\\
\hline
Train      & 1280830 & 1279170 & 2560000\\
Validation       & 319819 & 320181 & 640000\\
\hline\hline
\end{tabular}
\caption{Number of jets used for training and validating the top quark and non-top quark jet categories. When training the network on subsets of layers, only 198k jets were used for validation.}
\label{table:Njets}
\end{table}

We take reconstructed jets clustered using the anti-k$_{t}$ algorithm \cite{akt} with a radius parameter R of 0.8, or so-called AK8 jets, and require $\pt>400\GeV$ and $|\eta|<1.37$ for our event selection. Here, $\eta$ is the pseudorapidity and equates to the polar angle of the CMS detector according to $\eta = -\ln (\tan{\frac{\theta}{2}})$. This $\eta$ cut is to ensure that the jet image does not extend beyond the $|\eta|<2.4$ acceptance limit of the current CMS tracker.
Additionally, for the top jets we require the generator-level top quark, its bottom quark and $W$-boson daughters, and $W$-boson's daughters to be within an angular separation of $\Delta R=\sqrt{\Delta\eta^2+\Delta\phi^2}<0.8$ from the reconstructed AK8 jet axis, where $\phi$ is the azimuthal angle of the CMS detector.
In order to avoid biases caused by the different $\pT$ distributions of the jets, we pseudo-randomly drop jets from the three QCD samples such that the total number of jets and $\pT$ distribution of the $\ttbar$ sample is reproduced. After the $\pT$-resampling, the QCD and $t\bar{t}$ jets are split into the training and validation sets detailed in Table \ref{table:Njets}.

\section{CMS Detector, Images, and Network Training\label{sec:Images}}

CMS is a multi-purpose detector composed of several cylindrical subdetector layers, with both barrel and endcap sections, encasing a primary interaction point. It features a large $B=3.8$ T solenoid magnet to bend the trajectories of charged particles that aid in $\pT$ measurement \cite{tdr-magnet}. At the innermost layers there is a silicon tracker used to reconstruct the trajectory of charged particles and find their interaction vertices. The tracker can be divided in two parts the silicon pixel detector and silicon strip detector \cite{tdr-tracker}. The silicon pixel detector is the inner most part and, for the data taking years used in this study, composed of three layers in the barrel region (BPIX) and three disks in the endcap region (FPIX). Each layer is composed of pixel sensors that provide a precise position of the passage of a charged particle. The pixel detector provides crucial information for vertexing and track seeding. The outer part of the tracking system is composed of several layers of silicon strip sensors. 
This is followed by the electromagnetic calorimeter (ECAL) to measure the energy of electromagnetically interacting particles, then the hardonic calorimeter (HCAL) to measure the energy of hadrons \cite{tdr-ecal, tdr-hcal}. These are surrounded by the solenoid magnet which is finally encased by the muon chambers to detect the passage of muons \cite{tdr-muon}.

Following our previous work \cite{e2e1,e2e2}, we construct jet images using low-level detector information where each subdetector is projected onto an image layer, or several layers in the case of the tracker, in a grid of 125 x 125 pixels with the image centered around the most energetic HCAL deposit of the jet. Each pixel corresponds to the span of an ECAL barrel crystal which covers a 0.0174 $\times$ 0.0174 in the $\eta-\phi$ plane, giving our images an effective $\Delta R$ of 2.175. 
Reconstructed particle tracks are weighted by their reconstructed $\pT$ and their location is projected to an ECAL crystal.
To improve the identification of tracks coming from the hadronization of b quarks, we added additional layers motivated by the long flight distance of b hadrons producing reconstructed tracks that do not converge to the primary vertex. 
These new layers can be split into two categories: derived track quantities and reconstructed hits.

\begin{figure*}[htbp] 
\centering
\includegraphics[width=17.2cm]{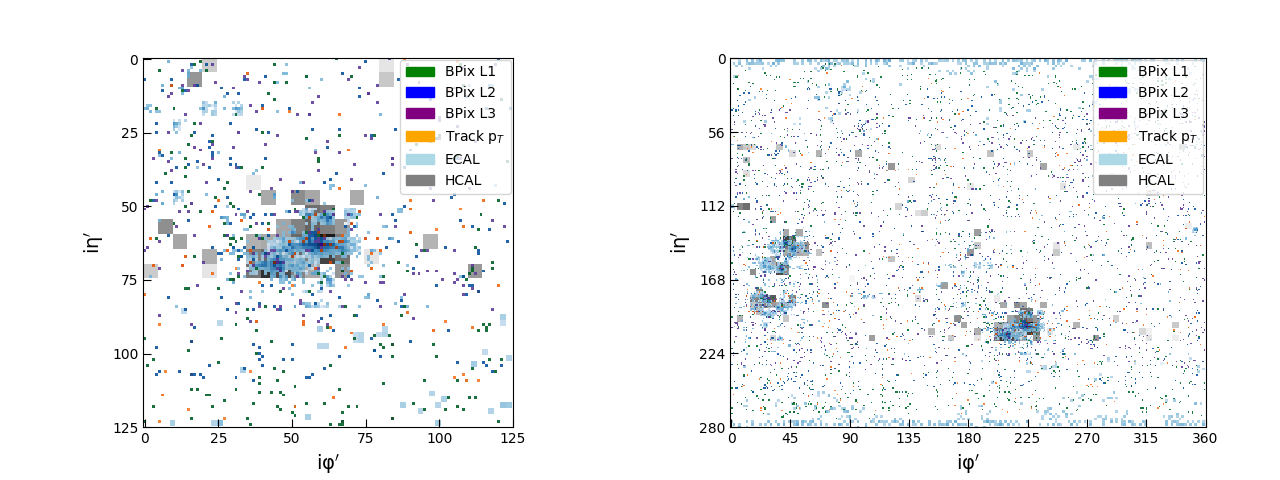}
\caption{Composite images of a simulated boosted top quark event. Images are produced for a single cropped jet image (left) and the full CMS detector (right).}
\label{fig:granularity}
\end{figure*}

We first add two new layers corresponding to the reconstructed tracks weighted by the transverse ($d0$) and longitudinal ($dZ$) components of their impact parameter (IP) significance.
The IP is defined as the distance vectors of minimum approach between the track helix and the primary vertex. To obtain the IP significance, the $d0$ and $dZ$ values are divided by their respective uncertainties. For this study, any $d0$ ($dZ$) values larger than 10 cm (20 cm) are suppressed to zero to prevent training degradation caused by the inclusion of tracks with a superfluously large IP. Such tracks are expected to originate from photon conversions in the tracker or from poor track reconstruction, and these cuts are not expected to negatively impact network performance.

\begin{table*}[ht] 
\centering
\begin{tabular}{l  c  c}
\hline\hline
\textbf{Layer Combinations} & \textbf{ROC-AUC} & \textbf{Sig-eff at 1\%}\\

    \hline
    Track $\pt$ (baseline) & 0.955$\pm$0.002 & 39.6\%\\
    Track $\pt$ + ECAL + HCAL (nominal) & 0.967$\pm$0.002 & 49.5\%\\
    Track $\pt$ + $d0$ + $dZ$ & 0.972$\pm$0.002 & 57.2\%\\
    Track $\pt$ + $d0$ + $dZ$ + ECAL + HCAL & 0.981$\pm$0.002 & 64.4\%\\
    \hline
    BPIX & 0.947$\pm$0.002 & 38.3\%\\
    BPIX\,+\,Track\,$\pt$ & 0.965$\pm$0.002 & 47.6\%\\
    BPIX\,+\,ECAL\,+\,HCAL & 0.975$\pm$0.002 & 56.9\%\\
    BPIX\,+\,Track\,$\pt$+$d0$+$dZ$ & 0.977$\pm$0.002 & 61.3\%\\
    \hline
    BPIX\,+\,Track $\pt$+$d0$+$dZ$\,+\,ECAL\,+\,HCAL & 0.9824$\pm$0.0013 & 66.41\%\\
    \hline
    Fully-Connected Network (groomed mass, substructure, track IPs) & 0.9258$\pm$0.0013 & 43.10\% \\
\hline\hline
\end{tabular}
\caption{Performance of the classifier trained up to 20 epochs. Results are shown for different combinations of tracking and calorimeter layers and evaluated on a sample of 198k jets.} 
\label{table:pt}
\end{table*}

In an effort to extract as much information as possible from the tracking subdetector, we include additional low-level detector information in the form of tracking hits, traditionally used in track reconstruction.
There are multiple steps in the conversion from charge clusters produced via charged particles passing through the tracker to fully reconstructed tracks. 

In this study, we consider the RecHit information from the three layers of the BPIX, but not from the FPIX or the silicon strip detector, as network inputs.
BPIX RecHits are obtained by first clustering nearby pixels of a given sensor which pass an adjustable charge threshold. A straight line fits the pixel cluster to center of the beam, and it's angle with the sensor surface is used to compute a hit location which is corrected for the Lorentz drift the charges experience before being read off the sensor. 
Given the hit location on the sensor and location of the sensor in the detector, the location of the RecHit is obtained.
An in depth explanation of the tracker RecHits and how they were calculated can be found in \cite{opendataml}.

For this study, the $\eta$ and $\phi$ position of the RecHit is re-calculated with respect to the primary vertex of the collision rather than the geometric center of the detector. This is done so that the $\eta$ and $\phi$ of the RecHits better match the $\eta$ and $\phi$ of their corresponding tracks when reaching the ECAL, which would otherwise deviate due to the pixel detectors closeness to the beamline.
From the RecHits three new image layers are produced, one for each layer of the BPIX, where each image pixel intensity is set to the number of RecHits per $\eta$--$\phi$ image resolution in the corresponding layer. 

After these additions, the images used in this study are composed of eight separate layers. Three for the BPIX RecHits, three for the tracks weighted by their $\pT$, $d0$, and $dZ$ values, and two for the calorimeters. Figure \ref{fig:granularity} shows an end-to-end image featuring all the image layers considered in this work for a single jet and the full detector. The only layers that cannot be seen are the track $d0$ and $dZ$ values due to their overlap with the track $\pT$ layer.

The network architecture and hyperparameters used in this work closely follow what was previously used in \cite{e2e1,e2e2}, making use of a ResNet-15 convolutional neural network \cite{resnet} trained with the ADAM optimizer \cite{adam}. The initial learning rate is $5\times 10^{-4}$ and is explicitly reduced by half every 10 epochs. We found training for 20 epochs to be sufficient for convergence. However, for our final network evaluations we used models trained for 40 epochs. The network was developed using the TensorFlow library \cite{tensorflow}.
The detailed description of the network training as well as a timing comparison between multiple GPU and TPU architectures can be found in the appendix of this manuscript.
We found that the TPUv3-8 trained the model roughly 35\% faster than an NVIDIA Tesla V100, but both the TPU and V100 had training speeds that were roughly 20 times faster than an NVIDIA Tesla P100.

To compare our results to a more traditional approach to jet identification, we additionally train a fully-connected network with eight layers and a total of 843,602 trainable parameters with binary cross entropy loss and ADAM optimizer \cite {adam}. A standard rectified linear unit (relu) activation function was used between each of the fully-connected layers, and the output was passed through the softmax function. The network was trained on the jet softdrop mass \cite{softdrop}, the ratio of 2-subjettiness to 3-subjettiness \cite{N-subjettiness}, and the $d0$ and $dz$ values of the tracks with the five highest IPs.
These variables were chosen because they are typically used to better identify jets originating from the hadronization of top and bottom quarks \cite{deepCSV}.

\section{Jet Identification Results\label{sec:jetID}}

Table \ref{table:pt} shows the area under the receiver operator curve (AUC) and signal efficiency at 1\% background misidentification rate for the end-to-end classifier using different combinations of track and calorimeter layers.
When training on a subset of layers the networks were trained for 20 epochs and evaluated on the 198k jet subset of the jets used for validation as described in Table \ref{table:Njets}. When training on the full combination of eight image layers and for the fully connected-network, the networks were trained for 40 epochs and evaluated on the full 640k jet validation sample.
A statistical uncertainty is obtained on the AUC score by inverting the square-root of the number of jets used to evaluate the network, and the signal efficiency is given for the AUC score central value.

Our previous end-to-end deep learning results showed that the Track $\pT$ layer gave the best single layer performance for jet discrimination \cite{e2e2}. Therefore, we choose track $\pT$ layer performance as a baseline for our models' performance. From Table \ref{table:pt}, we observe that the largest single-subdetector performance increase comes with the inclusion of the $d0$ and $dZ$ track information, leading to an AUC score improvement of 0.014--0.017. Comparing rows 2 and 3, the combination of track $\pT$, $d0$, and $dZ$ outperforms the nominal combination layers despite the fact that the $\pT$ + $d0$ + $dZ$ images are agnostic to neutral particles. This is in agreement with \cite{e2e2} where the tracks were observed as the most important feature for jet discrimination, and other jet tagging approaches that require the presence of a b-tagged subjet tagged using IP variables \cite{csv, toptag, deepCSV}.

Models trained only on BPIX produced a weaker performance than those trained using the track $\pT$ information.
However, we observe multiple improvements in network performance after combining BPIX with other layers.
When training the network on BPIX1--3, ECAL, and HCAL layers we find that it outperforms the nominal baseline in \cite{e2e2}, improving the AUC score by 0.008 and the signal efficiency from 49.5\% to 56.9\%.
Comparing rows 3 and 8 of Table \ref{table:pt} shows that adding the BPIX to the track $\pT$ + $d0$ + $dZ$ images improves the AUC by 0.005 and the signal efficiency from 57.2\% to 61.3\%. 

The bottom two rows of Table \ref{table:pt} show the performance of network trained on all 8 channels and a fully-connected network. The network trained on all 8 channels attains an AUC score of 0.9824$\pm$0.0013 and a signal efficiency of 66.41\% at 1\% misidentification. Comparing this to a fully-connected deep learning based top tagger, we find that all of the combination of layers explored outperform the fully-connected network's AUC score metric of $0.9258\pm0.0013$. 

\section{Interpretation and Discussion\label{sec:dicussion}}

An in depth look at the performance of neural networks trained on different layer combinations provides insight into which features the network is learning. The observation that the strongest single subdetector performance comes from the reconstructed tracks weighted by their $\pT$ and IP variables is in agreement with  expectations based on the current understanding of high momentum top jets. We expect a large number of high $\pT$ tracks, due to the jet containing three merged subjets, and a small subset of tracks having large IP values, attributed to a decaying B-meson. What is particularly interesting is that the network is able to successfully extract this IP information from the addition of the $d0$ and $dZ$ layers to the track $\pT$ image layer. 
The track-only images are composed of a set of perfectly overlapping sparse layers, and our initial intuition was that the 2D convolutional layers would have difficulty extracting new information due to its lack of new spatial structure.
We instead observe that this is not the case, and that these track-only images achieved an AUC of 0.972$\pm$0.002. This shows a large performance boost over our baseline comparison, and outperforms the nominal combination of layers that instead added new non-sparse spatial information to the baseline track $\pT$ images.

Further insight into the utility of IP variables can be observed by comparing a networks AUC performance to it's signal efficiency. When comparing performance between networks, we sometimes observe that the network with a higher AUC score does not always possess a higher signal efficiency as well.
As mentioned in Section \ref{sec:jetID}, we observe that the fully-connected network has a lower AUC score than any of the end-to-end networks, but that its signal efficiency is slightly larger than the networks trained only on track $\pt$ or BPIX1--3. We also observe this when comparing the network trained on the BPIX1--3, ECAL, and HCAL layers to the network trained on track $\pt$ + $d0$ + $dZ$. The BPIX1--3, ECAL, and HCAL network has a larger AUC score, but lower signal efficiency.
In both of these cases, we find that the networks with direct access to IP information have larger signal efficiencies at 1\% background rejection, but still have a lower AUC value.
Further investigation into the Receiver Operator Curves showed that the addition of IP variables primarily improves the networks background rejection at discriminator values with very high signal efficiency greater than 90\%. However, this has a smaller impact on the networks background rejection at discriminator values with lower signal efficiency.


The second insight comes from the performance of the BPIX RecHits. As mentioned in Section \ref{sec:jetID}, the BPIX  do not show a strong standalone single sub-detector performance. However, this is to be expected for multiple reasons. The pixel detector has an $\eta$ and $\phi$ resolution of 10 $\mu$m, giving the inner most layers a 1D spatial resolution that is almost eight times finer than the ECAL \cite{tdr-tracker, tdr-ecal}. Furthermore, we only considered the barrel region of the pixel detector, and do not include any RecHits from the forward region of the pixel detector. Any jets that border the $\eta$ acceptance of this study will be missing RecHits from portions of the BPIX layers.
Finally, our network is agnostic to each layer's distance from the beamline, giving the network incomplete information about the RecHits global positioning. For example, the RecHits will drift in $\phi$ as the charged particle bends in the CMS detector's magnetic field. But unless more layers are added to the image, the network does not have enough information to know the order of each hit nor the direction in $\phi$ the particle is moving. But despite the shortcomings of our current RecHit implementation, we find remarkable results. With the exception of the final layer combination, where BPIX RecHits are added to images composed of track $\pT$ + $d0$ + $dZ$ + ECAL + HCAL information, we note that adding the BPIX RecHits gives a statistically significant increase in network performance. The most notable are cases where BPIX RecHits are added on top of the tracking variables (1), and the case where BPIX RecHits are used in lieu of the derived tracking information (2).

In the first case (1), we see that the network is able to use the BPIX to exploit jet features which it could not parse from the derived track quantities alone. One possible feature is the track charge, where motion through $\phi$ can be combined with the final location of the track to determine its direction of curvature of the track. However, more abstract features may also exist in these images. In the case of (2), the network does not use any reconstructed variables for its inputs. We see that despite the lack of derived variables, the network outperforms the track $\pT$ + $d0$ + $dZ$ images, and only performs marginally worse than the final performance on the full images.
The overall success of our network's ability to learn from BPIX RecHits paves the foundation for future studies of end-to-end taggers where no derived variables are used.

\section{Conclusions\label{sec:Conclusions}}

In this work we have extended the end-to-end deep learning technique to top quark jet classification. To enhance the performance of the classifier we added additional layers containing information about track parameters and pixel detector reconstructed hits, marking the first top-tagging algorithm which uses tracking RecHits as input variables. The model was trained using CMS Open Data datasets containing low-level tracking information \cite{ttbarod,qcdod1,qcdod2,qcdod3}.

The end-to-end classifier trained on all input features achieves an AUC performance of 0.9824$\pm$0.0013. 
We find that the addition of $d0$ and $dZ$ variables gives the largest boost to network performance when compared to subdetector information used in previous end-to-end jet discrimination studies \cite{e2e2}. At ECAL granularity, BPIX RecHits do not provide the network with information that is not present in the combination of track $pT$, $d0$, $dZ$, ECAL, and HCAL layers. However, we find that it still improves subgroups of these layers, and the network achieves an AUC score of 0.975$\pm$0.002 when training on images void of derived variables. These findings lay the ground work for future studies which look to incorporate RecHits from the full CMS tracker, higher-resolution training, and to explore new deep learning architectures that can fully exploit the tracker granularity.


\section*{\label{sec:Acknowledgments}Acknowledgments}
We would like to thank the CMS Collaboration and the CERN Open Data group for releasing their simulated data under an open access policy. We strongly support initiatives to provide such high-quality simulated datasets that can encourage the development of novel but also realistic algorithms, especially in the area of machine learning. We believe their continued availability will be of great benefit to the high energy physics community in the long run.

We would also like to thank Google for access to Google Cloud and TPU computing nodes which helped us speed up the training of machine learning models used in this study.

\pagebreak
\appendix


\section{Timing Performance Comparison\label{sec:Tpu}}

A number of factors affect the training speeds of the classifiers and their memory requirements:
The number of events used during training, the number of image layers in each event, and the resolution of each of the images.
Compared to our previous work \cite{e2e2}, the number of training events has increased by a factor of 3.2 and the number of image layers has increased by a factor 2.7.
For a single 125 x 125 pixel image, we can see that the size of an uncompressed image has increased from 183 kB in \cite{e2e2} to 488 kB.
These factors combined lead to a significant computational cost increase for training the network, both in training time and memory requirements. For this task training was carried out on multiple accelerated hardware architectures as a benchmark to compare their performance. The comparison was performed on two different graphical processing units (GPU) and a tensor processor unit (TPU), whose specifications are summarized in Table \ref{table:specs}. Additionally, a test was performed where we trained a single network on multiple V100 GPUs in parallel to observe how the training speeds scaled with number of processing units.
A discussion on the differences of the architectures and a detailed description on the external input/output (I/O) pipeline used when utilizing each architecture is described below.

\subsection{Architectures and I/O Pipeline}

The NVIDIA Tesla P100 is a GPU that utilizes the Pascal architecture \cite{tesla}. 
The Tesla P100 GPU was accessed on a shared cluster at the Fermilab National Accelerator Laboratory LHC Physics Center via a dedicated GPU worker node. The node accessed the GPU through a 12 GB/s PCIe connection using the CUDA Toolkit v9.1 drivers.
During training, data was stored and read from an HGST 1W10002 hard disc drive \cite{hdd} located on the GPU machine. Images were uncompressed, pre-processed, and sent to the GPU using a single Intel(R) Xeon(R) Silver 4110 8-core CPU \cite{intel-silver}.

The NVIDIA Tesla V100 uses the Volta architecture, incorporating eight tensor cores and an all-around higher performance than the Pascal architecture \cite{volta}. Unlike the P100, the V100 is able to make use of mixed precision operations, which were utilized for this comparison, to drastically increase the number of floating point operations per second (FLOPS).
During training, images were read from a solid state drives with better random I/O operations per second than traditional disc drives.
The computing node ran Cuda v11.0.2 drivers and used four Intel(R) Xeon(R) Gold 5118 12-core CPUs \cite{intel-gold} to perform data loading and pre-processing tasks, which were parallelized using the NVIDIA DGX-1 architecture \cite{nvidia-dgx}. Of the 48 available CPU cores, 20 were used for data loading.
When training the network on multiple GPUs we chose to train on one, two, four, and eight V100s using the the Horovod framework \cite{sergeev2018horovod}. For the scaling tests a larger batch size of 1024 tf examples was used.

\begin{table*}[t]
    \centering
    \begin{tabular}{l   c   c   c   c}
    \hline\hline
         \textbf{Processor} & \textbf{Manufacturer} & \textbf{Year Released} & \textbf{HBM Memory} & \textbf{Performance}\\ \hline
         Tesla P100 & NVIDIA & 2016 & 16 GB & 9.3 Single-Precision TeraFLOPS \\
         Tesla V100 & NVIDIA & 2017 & 32 GB / 16 GB  & 125 Mixed-Precision TeraFLOPS \\
         TPUv3-8 & Google & 2018 & 128 GB & 420 Mixed-Precision TeraFLOPS \\
         \hline
    \end{tabular}
    \caption{Comparison of architecture specification for the NVIDIA Tesla P100 \cite{tesla}, NVIDIA Tesla V100 \cite{volta}, and Google TPUv3-8 \cite{tpuv3} architectures. The floating point operations per second speed (FLOPS) are quoted based on the data type and architecture setup used during training.}
    \label{table:specs}
\end{table*}

The TPU is a type of AI-accelerated architecture designed by Google with the purpose of training and running inference on machine learning models \cite{tpu}, and can be accessed through the Google Cloud platform \cite{google-cloud}.
TPUs boast a high number of FLOPS, made possible by dedicated matrix multiplication units which make use of the bfloat16 data type \cite{bfloat}. Unlike GPU based architectures, the CPU cores used to fetch and pre-process batches all live on TPU board creating a low latency I/O pipeline which does not have to compete for CPU resources.
For this work, a TPUv3-8 running TPU software v1.14 was run using a type n1-standard-1 Google Cloud virtual machine node. Data used to train the networks was stored in Google Cloud buckets located in the same region as the virtual machine and TPU nodes to decrease time associated with transferring data during the training. This gives the cloud storage buckets comparable performance to persistent HDs in the local VM storage area \cite{bucket-storage}.

\subsection{GPU Scaling Performance}

\begin{figure}[b]
\centering
\includegraphics[width=8.6cm]{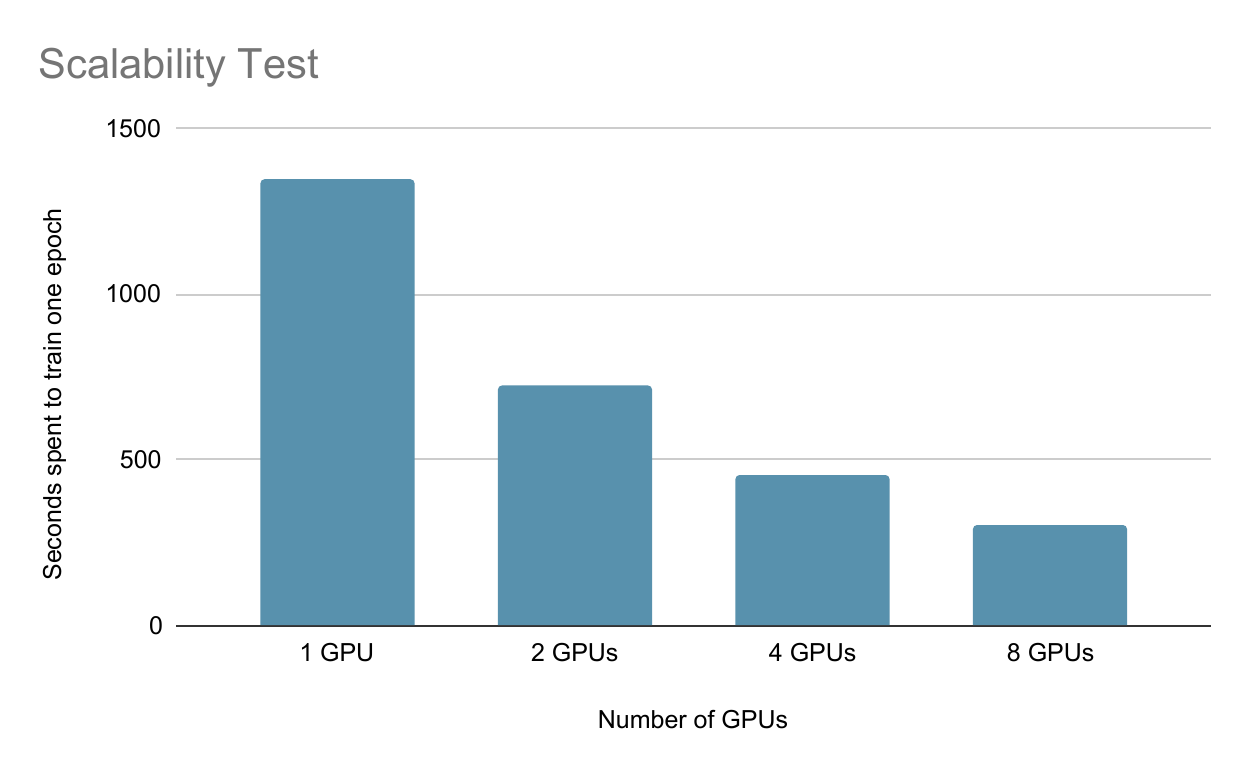}
\caption{Scaling end-to-end deep learning training on multiple GPUs.}
\label{fig:scaleGraph}
\end{figure} 

The Horovod framework \cite{sergeev2018horovod} was used to perform the GPU scaling, as it provides flexibility of scaling the training of the network to multiple GPUs. 
Horovod takes advantage of the inter-GPU and inter-node communication methods such as NCCL (Nvidia Collective Communications Library) and MPI (Message Passing Interface) to distribute the deep learning model parameters between various workers and aggregate them accordingly. 

Figure \ref{fig:scaleGraph} shows the training time when the final optimised model was scaled to multiple GPUs. Going from one to two GPUs, we observe a roughly 50\% decrease in training speeds. However, as we continue to double the GPUs, we observe diminishing returns in improvement. This is primarily due to the underlying input bottlenecks which are not addressed by improving computation parallelization. Additionally, adding more GPUs increases the output latency associated with averaging gradients.

\subsection{Computing Architecture Comparison}

\begin{figure}[t]
\centering
\includegraphics[width=8.6cm]{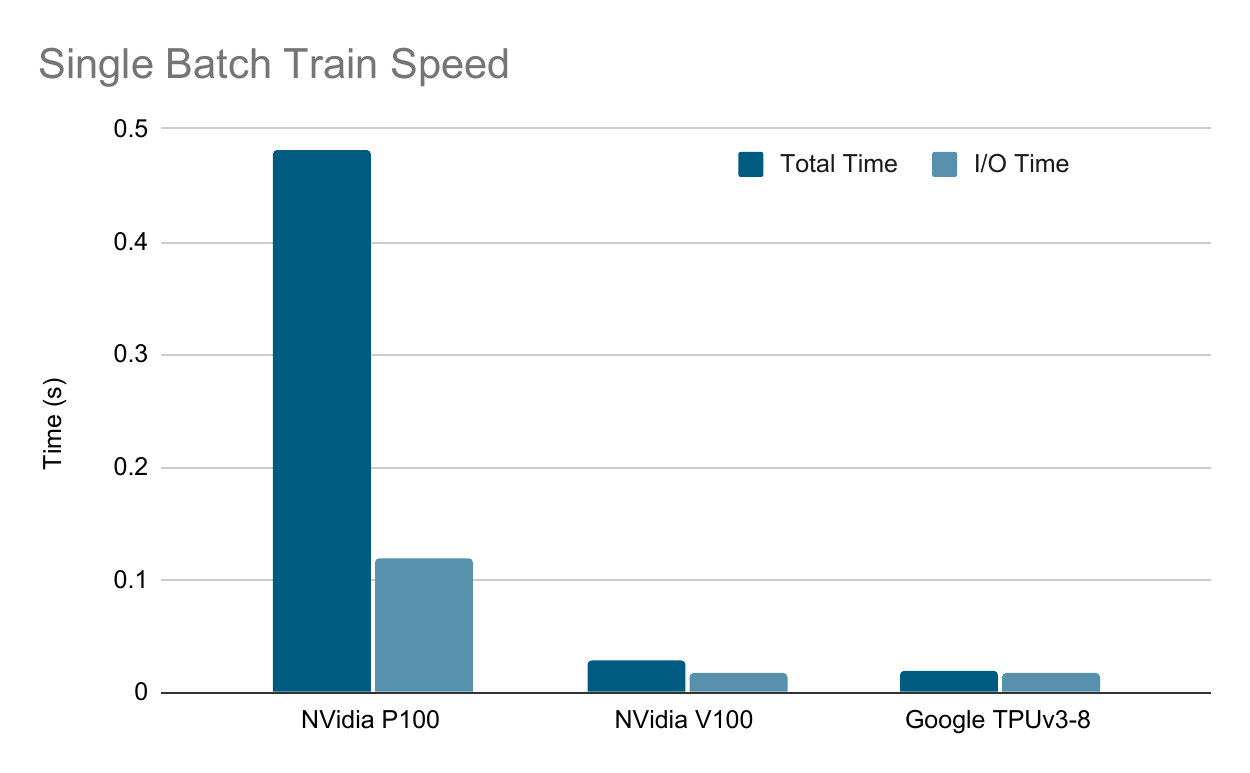}
\caption{Comparison of time taken to train over a single batch on the GPU and TPU architectures.}
\label{fig:timeGraph}
\end{figure}

Table \ref{table:tpu} and Figure \ref{fig:timeGraph} provide a timing comparison for training machine learning models using the three different architectures. The training performed on the P100 were drastically longer, partially stemming from low I/O speeds caused by the inefficient random reads associated with disk HDs \cite{random-sample} as well the weaker CPU used for building batches and sending them to the GPU. When utilizing the Tesla P100, we observe that improvements to the computing clusters I/O infrastructure could decrease the required time to train on a single epoch by up to 2.6 hours.

The Tesla V100 and TPUv3-8 give much stronger performance and were accessed utilizing I/O pipelines with comparable performance. However, in both cases we see that over half of the time associated with training a batch is spent on I/O. By subtracting the single batch I/O time from the single batch train time, we obtain an approximate computation time.
From this, we see that the TPUv3-8 spent approximately 2.6 ms to perform forwards and backwards propagation calculations, which is a factor of four faster than the 11 ms required by the Tesla V100.
\\

\begin{table}[ht]
\centering
\begin{tabular}{l | c c c}
\hline\hline
\textbf{Category} & \textbf{Tesla P100} & \textbf{Tesla V100} & \textbf{TPUv3-8} \\
\hline
Training Batch Size & 32 jets & 64 jets & 64 jets \\
Batches Per Epoch & 80k & 40k & 40k \\ \hline
I/O Time (batch) & 0.119 s & 0.0180 s & 0.0176 s\\
Train Time (batch) & 0.481 s & 0.0290 s & 0.0202 s\\
Train Time (epoch) & 321 min & 19 min & 14 min\\ 
\hline\hline
\end{tabular}
\caption{Comparison of I/O and training time for different computing architectures. A larger batch size was used when training on the Tesla V100 and TPUv3-8. Training times were found to vary by approximately 10\% between epochs.}
\label{table:tpu}
\end{table}

\subsection{Timing Conclusion}

Identical copies of the model were trained on a single NVIDIA Tesla P100 GPU, a single NVIDIA Tesla V100 GPU, and a TPUv3-8 accessed through the Google Cloud platform.
We observe that one of the largest training speed improvements come from the utilization of a more optimized I/O infrastructures. For many existing computing clusters, improvements to the I/O pipeline could serve as a relatively cheap way to greatly improve training speeds.

We find that both the Tesla V100 GPU and the TPUv3-8 offer a significant training time improvement over a Tesla P100 GPU when training on end-to-end jet images, but that the TPUv3-8 still maintained faster computation times.
We also trained a identical copies of the network in parallel on multiple NVIDIA Tesla V100 GPUs to observe the relationship between number of GPUs and training speeds. From this, we were able to see a roughly linear relationship between the number of GPUs utilized and the time it took to train the network. However, at high number of GPUs a plateau starts to occur as I/O becomes the performance bottleneck. 



\bibliographystyle{lucas_unsrt}
\bibliography{bibliography}{}

\end{document}